\documentclass[aps,prc,twocolumn,superscriptaddress,showpacs,showkeys,
               floatfix,nofootinbib,altaffilletter]{revtex4}

\usepackage{xspace}
\usepackage{color}\definecolor{darkblue}{rgb}{0,0,0.5}
\usepackage{hyperref}
\hypersetup{
    bookmarks=true,         % show bookmarks bar?
    unicode=false,          % non-Latin characters in Acrobat's bookmarks
    pdftoolbar=true,        % show Acrobat's toolbar?
    pdfmenubar=true,        % show Acrobat's menu?
    pdffitwindow=true,      % page fit to window when opened
    pdftitle={My title},    % title
    pdfauthor={Author},     % author
    pdfsubject={Subject},   % subject of the document
    pdfnewwindow=true,      % links in new window
    pdfkeywords={keywords}, % list of keywords
    colorlinks=true,       % false: boxed links; true: colored links
    linkcolor=darkblue,          % color of internal links
    citecolor=darkblue,        % color of links to bibliography
    filecolor=magenta,      % color of file links
    urlcolor=darkblue,         % color of external links
}

\newcommand{\FlukaLong}{{\scshape Fluka2008}\xspace}
\newcommand{\Fluka}{{\scshape Fluka}\xspace}
\newcommand{\UrqmdLong}{{\scshape U}r{\scshape qmd1.3.1}\xspace}
\newcommand{\Urqmd}{{\scshape U}r{\scshape qmd}\xspace}

\newcommand{\Corsika}{{\scshape Corsika}\xspace}
\newcommand{\Venus}{{\scshape Venus}\xspace}
\newcommand{\VenusLong}{{\scshape Venus4.12}\xspace}

\usepackage[latin2]{inputenc}
\usepackage{indentfirst}
\usepackage{enumerate}

\usepackage{amsmath}
\usepackage{amssymb}
\usepackage[english]{babel}
\usepackage{url}

\newif\ifpdf

\ifx\pdfoutput\undefined

\pdffalse

\else

\pdfoutput=1

\pdftrue

\fi

\ifpdf

\usepackage[pdftex]{graphicx}

\else

\usepackage{graphicx}

\fi

\usepackage{stmaryrd}

\begin{document}
\newcommand{\ra}[1]{\renewcommand{\arraystretch}{#1}}
\DeclareGraphicsExtensions{.pdf,.png,.eps,.jpg,.ps}
%\DeclareCaptionStyle
\cleardoublepage

\pagenumbering{roman}

\title{\large \bf
  Measurement of Production Properties
  of Positively Charged Kaons \\
  in Proton--Carbon Interactions at 31~GeV/$c$}

\begin{abstract}
  Spectra of positively charged kaons in p+C interactions at 31~GeV/$c$ were
  measured with the NA61/SHINE spectrometer at the CERN SPS. The
  analysis is based on the full set of data collected in 2007 with a
  graphite target with a thickness of 4\% of a nuclear interaction
  length. Interaction cross sections and charged pion spectra
  were already measured using
  the same set of data. These new measurements in
  combination with the published
  ones are required to improve predictions of the neutrino flux for
  the T2K long baseline neutrino oscillation experiment in Japan. In
  particular, the knowledge of kaon production is crucial for
  precisely predicting the intrinsic electron neutrino component and 
  the high energy tail of the T2K beam. 
  The results are presented as a function of laboratory
  momentum in 2 intervals of the laboratory polar angle covering the
  range from 20 up to 240~mrad.  The kaon spectra are compared with
  predictions of several hadron production models. Using the
  published pion results and the new kaon data, the $K^+$/$\pi^+$ ratios are computed.
\end{abstract}

\clearpage

 % Authors in alphabetical order.
 \author{N.~Abgrall}\affiliation{University of Geneva, Geneva, Switzerland}
 \author{A.~Aduszkiewicz}\affiliation{Faculty of Physics, University of Warsaw, Warsaw, Poland}
 \author{T.~Anticic}\affiliation{Rudjer Boskovic Institute, Zagreb, Croatia}
 \author{N.~Antoniou}\affiliation{University of Athens, Athens, Greece}
 \author{J.~Argyriades}\affiliation{University of Geneva, Geneva, Switzerland}
 \author{B.~Baatar}\affiliation{Joint Institute for Nuclear Research, Dubna, Russia}
 \author{A.~Blondel}\affiliation{University of Geneva, Geneva, Switzerland}
 \author{J.~Blumer}\affiliation{Karlsruhe Institute of Technology, Karlsruhe, Germany}
 \author{M.~Bogusz}\affiliation{Warsaw University of Technology, Warsaw, Poland}
 \author{L.~Boldizsar}\affiliation{KFKI Research Institute for Particle and Nuclear Physics, Budapest, Hungary}
% ...

 \collaboration{\bf The NA61/SHINE Collaboration}
 \noaffiliation

 \date{\today}
 \pacs{13.85.Lg,13.85.Hd,13.85.Ni}
 \keywords{p+C interaction, inclusive kaon spectra}

 \maketitle
 \clearpage

 %\tableofcontents
 %\listoftables
 %\listoffigure

 % \pagestyle{fancyplain}

 \pagenumbering{arabic}

 \section{Introduction}

 The SPS Heavy Ion and Neutrino Experiment (NA61/SHINE) at CERN pursues
 a rich physics program~\cite{proposala,add1,proposalb,Status_Report_2008}.
 Hadron production
 measurements in p+C and $\pi$+C interactions 
will improve calculations of neutrino fluxes in the T2K
 experiment~\cite{T2K}, and simulations of cosmic-ray air showers in
 the Pierre Auger and KASCADE experiments~\cite{Auger,KASCADE}. The
 heavy ion program investigates p+p, p+Pb and nucleus+nucleus collisions
 at SPS energies, to study the onset of deconfinement and search for
 the critical point of strongly interacting matter. Charged pion
 spectra in p+C interactions at 31~GeV/$c$ were already
 published~\cite{pion_paper} and used for neutrino flux prediction
 in T2K~\cite{T2K1stnue}. This article presents new measurements of
 positively charged kaon spectra in p+C interactions at 31 GeV/$c$,
 based on the data collected during the first running period in
 2007. A detailed description of the experimental apparatus and
 analysis techniques can be found in~\cite{pion_paper}.

 T2K -- the long baseline neutrino experiment from J-PARC in Tokai to
 Kamioka (Japan) -- aims to precisely measure
 the $\nu_{\mu} \rightarrow\nu_e$ appearance and
 $\nu_{\mu}$ disappearance~\cite{T2K,T2K1stnue}.
 The neutrino beam is generated by the J-PARC
 high intensity 30~GeV (kinetic energy) proton beam interacting in a
 90~cm long graphite target to produce $\pi$ and K mesons, which decay
 emitting neutrinos. The resulting neutrino beam is aimed towards a near
 detector complex, 280~m from the target, and to the Super-Kamiokande
 (SK) far detector located 295~km away at 2.5~degrees off-axis from
 the 
 hadron beam. Neutrino oscillations are probed by comparing
 the neutrino event rates measured in SK to the predictions of a
 Monte-Carlo simulation based on flux calculations and near detector
 measurements. Until the NA61/SHINE data were available, these flux
 calculations were based on hadron production models tuned to sparse
 available data, resulting in systematic uncertainties which are large
 and difficult to evaluate. Direct measurement of particle production
 rates in p+C interactions allows more precise and
 reliable estimates. Presently, the T2K neutrino beam-line is set up
 to focus positively charged hadrons, to produce a $\nu_{\mu}$
 beam. While charged pions generate most of the low energy neutrinos,
 positively charged kaons generate the high energy tail of the T2K
 beam, and contribute substantially to the intrinsic $\nu_{e}$
 component in the T2K beam.

\begin{figure}[h]
   \begin{center}
    \includegraphics[width=1.\linewidth,height=0.3\textheight]{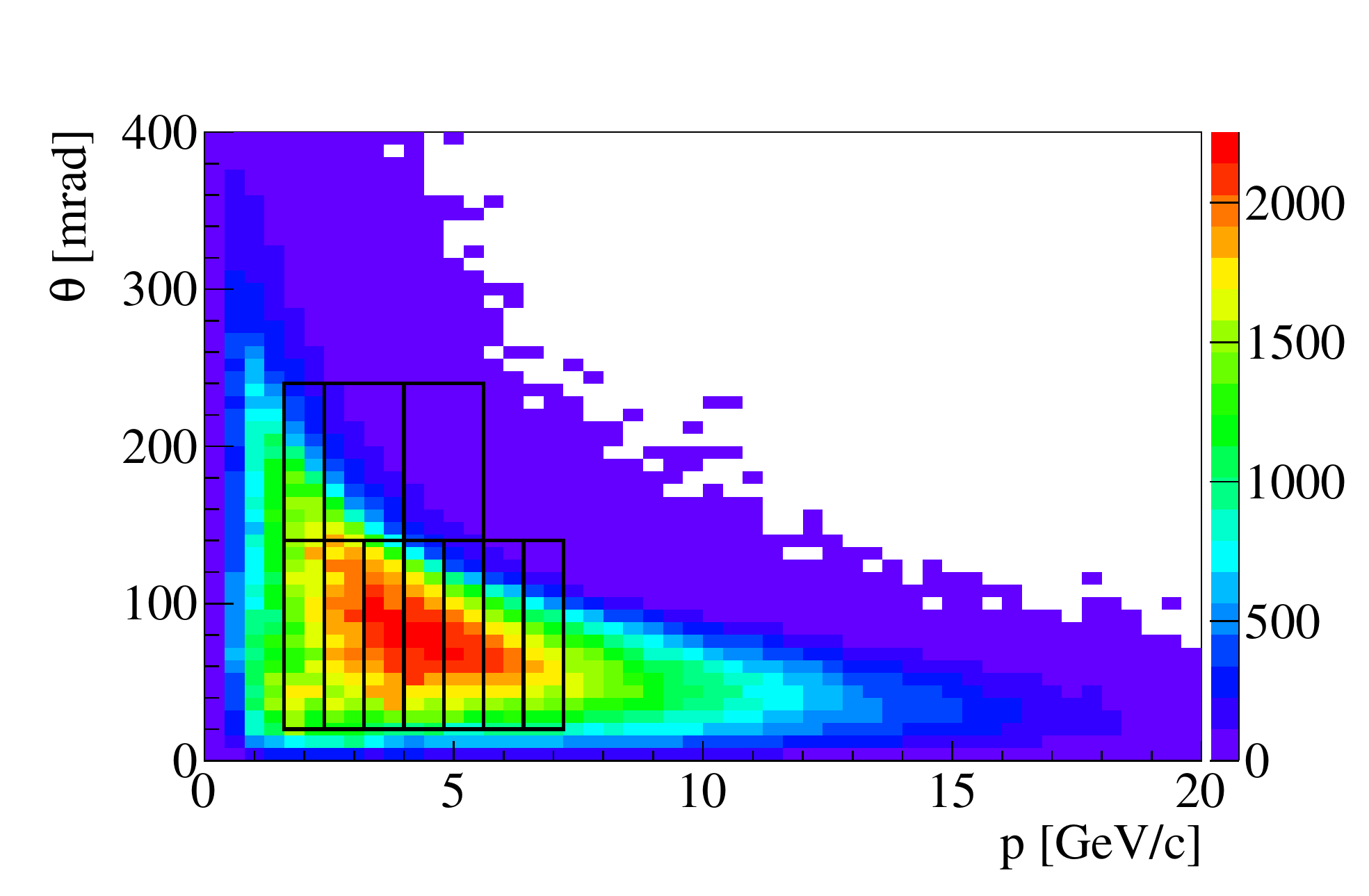}
  \end{center}
  \vspace*{-0.5cm}
  \caption{ (Color online) The prediction from the T2K beam
    simulation; the $\{ p,\theta\}$ distribution for positively
    charged kaons weighted by the probability that their decay
    produces a neutrino passing through the SK detector. The binning
    used in the present analysis is superimposed; the kinematic range
    considered is 1.6$<\!p\!<$7.2~GeV/$c$ and
    20$<\!\theta\!<$240~mrad.  }
    \label{fig:p-theta-k+SK}
\end{figure}

 Positively charged kaons whose daughter neutrinos pass through the SK
 detector constitute the kinematic region of interest, shown in
 Fig.~\ref{fig:p-theta-k+SK} in the kinematic variables $p$ and
 $\theta$ -- the momentum and polar angle of particles in the
 laboratory frame. The low statistics available in the 2007 pilot data
 set imposes a ${p,\theta}$ binning which covers only the most
 populated region of phase space relevant for T2K.  Moreover, the
 statistics of the 2007 data does not allow for measurements of
 negatively charged kaons.  An order of magnitude larger data set was
 recorded in 2009, and, when analyzed, will lead to essentially full
 coverage. The NA61/SHINE data on kaon production will allow also to
 test and improve existing hadron production models in an energy
 region which is not well constrained by measurements at present.
 Several $K^+$ production measurements in this energy range were
 performed
 previously~\cite{Abbott,Aleshin,Allaby,Dekkers,Eichten,Marmer,Vorontsov}.

\section{The experimental setup} \label{sec:setup}

The NA61/SHINE apparatus (for details see
Ref.~\cite{pion_paper}-Sec.~II) is a wide acceptance spectrometer made
of four large volume Time Projection Chambers (TPCs): two Vertex TPCs
(VTPC-1 and VTPC-2) placed in the magnetic field produced by two
superconducting dipole magnets and two Main-TPCs (MTPC) located
downstream symmetrically with respect to the beam line
(Fig.~\ref{fig:evt_disp}).  In the forward region, the setup is
complemented by a time-of-flight (\mbox{ToF-F}) detector array
horizontally segmented into 64 scintillator bars read out at both ends
by photo-multipliers.  The time resolution of each scintillator is
about 115~ps \cite{Sebastien}.  For the study presented here the
magnetic field was set to a bending power of 1.14 Tm.  This leads to a
momentum resolution $\sigma(p)/p^2$ in the track reconstruction of
about $5 \times 10^{-3}$~(GeV/$c$)$^{-1}$.  The target used is an
isotropic graphite sample with a thickness along the beam axis of 2
cm, equivalent to about 4\% of a nuclear interaction length,
$\lambda_{\mathrm{I}}$.  During the data taking the target was placed
80 cm upstream of the VTPC-1.
\section{Analysis technique}\label{Sec:ana}

This section presents details on data selection and binning,
the kaon identification method as well as corrections and
systematic errors.

\subsection{Data binning}\label{Sec:binning}
The adopted binning scheme is mainly driven by the available statistics
and is presented in Fig.~\ref{fig:p-theta-k+SK}. Details
can be found in Table~\ref{tab:k_results}.
The  highest $\theta$ limit
is determined by the requirement for the track to be in the
geometrical acceptance of the ToF-F detector.

Only two angular intervals are defined.  The lowest $\theta$ value is
set to 20 mrad in order to exclude tracks passing close to the edges
of the TPCs where the reconstruction efficiency is lower and the
calculation of the correction for the acceptance is less reliable.
The first angular interval extends up to 140 mrad so as to cover most
of the T2K relevant $\theta$ range and, combined with a 0.8 GeV/$c$
momentum bin width, to have a few thousands of selected tracks per
interval (Table \ref{tab:k_results}).  Measurements were performed up
to 7.2 GeV/$c$.  This choice comes from the fact that in the
relativistic rise region (above 4-5 GeV/$c$) particle identification
requires extracting the rapidly decreasing kaon signal from the
predominant proton one.  With the available statistics of 2007 data
the applied procedure turned out to be robust only up to about
7~GeV/$c$.

\subsection{Event and track selection}\label{sec:track_sel}
\begin{figure}[ht]
  \begin{center}
    \includegraphics[width=0.83\linewidth,height=0.37\textheight,angle=-90]{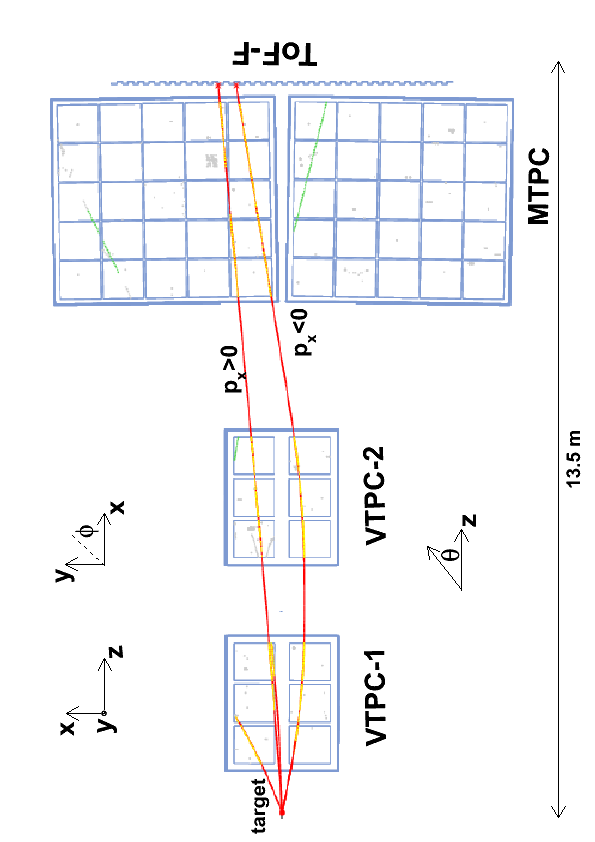}
  \end{center}
  \caption{\label{fig:evt_disp} (Color online) Schematic drawing of
    the experimental apparatus (Sec.~\ref{sec:setup}) and a
    reconstructed p+C interaction event.  Yellow (green) points
    indicate TPC clusters (not) associated to reconstructed primary
    tracks.  Stars correspond to hits reconstructed in the ToF-F.  Red
    lines are the fitted particle trajectories.  }
\end{figure}

This analysis is based on 452$\times 10^3$ reconstructed events
collected during the 2007 data taking. Only events for which a beam
track is properly reconstructed are selected. The beam trajectory is
measured with a set of Beam Position Detectors (BPD) placed upstream
of the target (\cite{pion_paper}-Sec. III,V).  Several criteria were
applied to select well-measured positively charged tracks in the TPCs
and ensure high reconstruction efficiency as well as to reduce the
contamination of tracks from secondary interactions:
\begin{enumerate}[(i)]
\setlength{\itemsep}{1pt}
\item  track momentum fit at the
interaction vertex should have converged,
\item a minimum of 12 reconstructed points in the two TPCs used for
  momentum measurement, VTPC-1 and VTPC-2, is required,
\item the distance of closest approach of the fitted track to the
  interaction point (impact parameter) is required to be smaller than
  4~cm in both transverse directions,
\item the track must leave the primary vertex at an azimuthal angle
  $\phi$ within $\pm$$20^{\circ}$ around the horizontal plane, for the
  first $\theta$ interval, and $\pm$$10^{\circ}$ for the second; this
  excludes most of the tracks traversing the detector in the regions
  where the reconstruction capability is limited by the magnet
  aperture or by the presence of uninstrumented regions in the
  VTPCs, 
\item the track must have an associated ToF-F hit.
\end{enumerate}
The position of a ToF-F hit is determined only in the $x$ direction
and with a precision given by the width of the scintillator slat
producing the signal ($\sim\!10$ cm) (Fig.~\ref{fig:evt_disp}).  A
ToF-F hit is then associated to a track if the trajectory can be
extrapolated to the pertaining slat.

The relatively short proper decay length of  kaons ($c\tau\!\sim\!3.7$ m) compared to the
longitudinal extension of the detector ($\sim\!13$ m) implies that
almost 60$\%$ of these particles produced in the target
at the lowest laboratory momentum considered,~1.6 GeV/$c$,
decay before reaching the ToF-F array.

Thanks to the high $Q$ values of the kaon decay channels,
 kink topologies
are correctly reconstructed as a primary and a secondary track
(i.e. not fitted to the primary vertex)
 with an  efficiency higher than 98$\%$.
 Nevertheless, a ToF-F hit would still be associated to a kaon
 decaying in flight if the secondary particle produces a hit along the
 same scintillator bar in which a hit from the primary is expected. As
 a consequence the time of flight measurement could be significantly
 biased. Such special topologies can be effectively rejected by
 considering only tracks reconstructed until the downstream edge of
 the MTPCs.  (see Sec. \ref{Sec:corrections}).  This can be achieved
 by the following cut: \vspace{-0.1cm}
\begin{enumerate}
\item [(vi)]
the $z$ position of the last reconstructed point, $z_{last}$,
must differ by less then 50 cm from the position of the last potential point
(see also Fig.~\ref{fig:evt_disp}).
\end{enumerate}
The sample can be divided into tracks emitted in the same direction in
which they are bent by the magnetic field (i.e. $p_x\!>\!0$ for
positive charges) and tracks emitted in the opposite direction (see
Fig.~\ref{fig:evt_disp}).  These two independent sub-samples were
treated separately.  In fact, for the same $\{p,\theta\}$ bin the
fraction of accepted tracks can differ significantly between the two
topologies. In order to improve the accuracy of the Monte Carlo
correction and to reduce the dependence on the model used for
simulation (Sec.~\ref{Sec:corrections}),
only one track topology per bin, the one with the highest acceptance, is chosen.\\

\subsection{Combined particle identification} \label{Sec:combined}

\begin{figure*}[htb]
  \begin{center}
    \includegraphics[width=0.32\linewidth,scale=0.4]{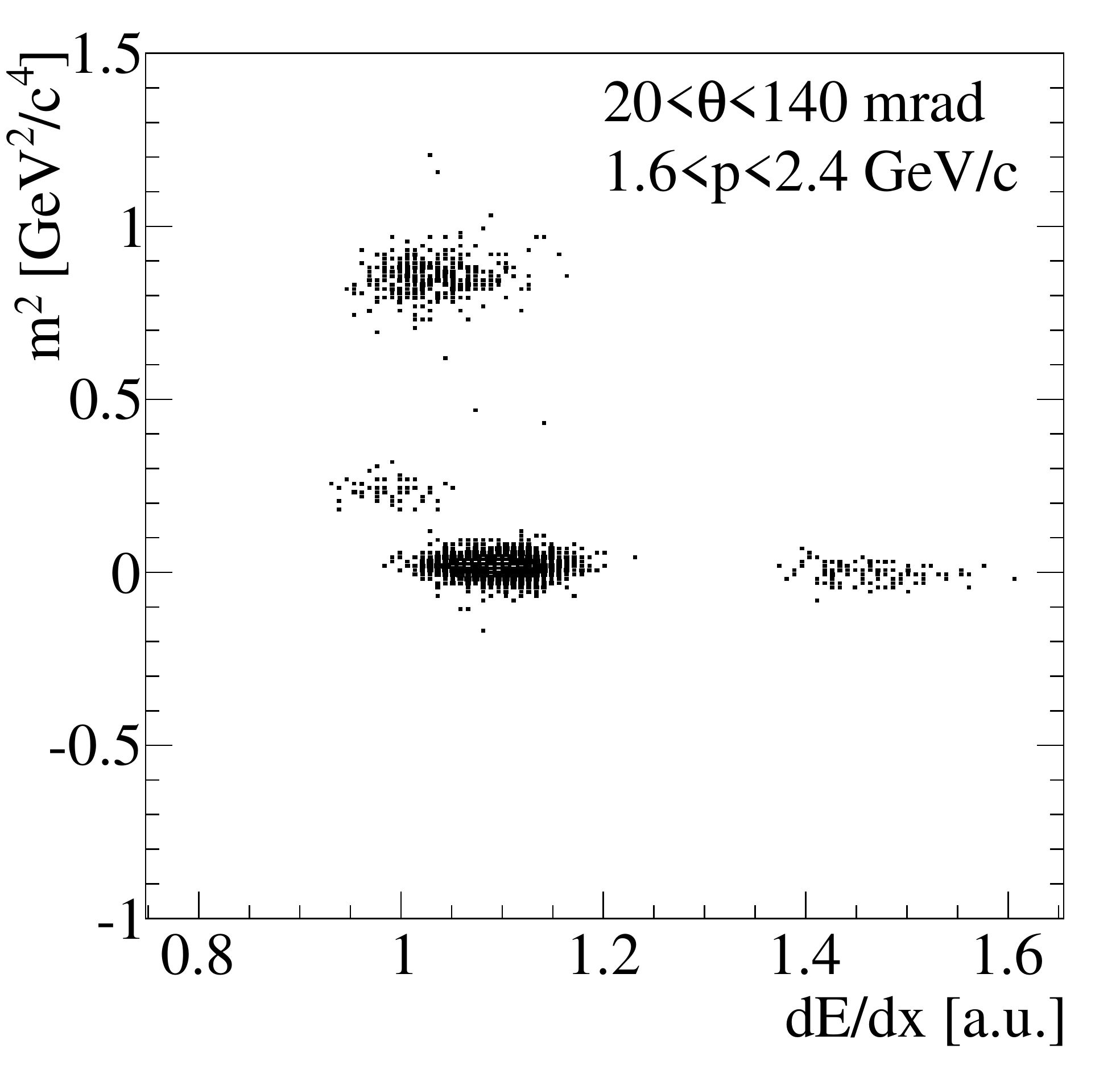}
    \includegraphics[width=0.32\linewidth,scale=0.4]{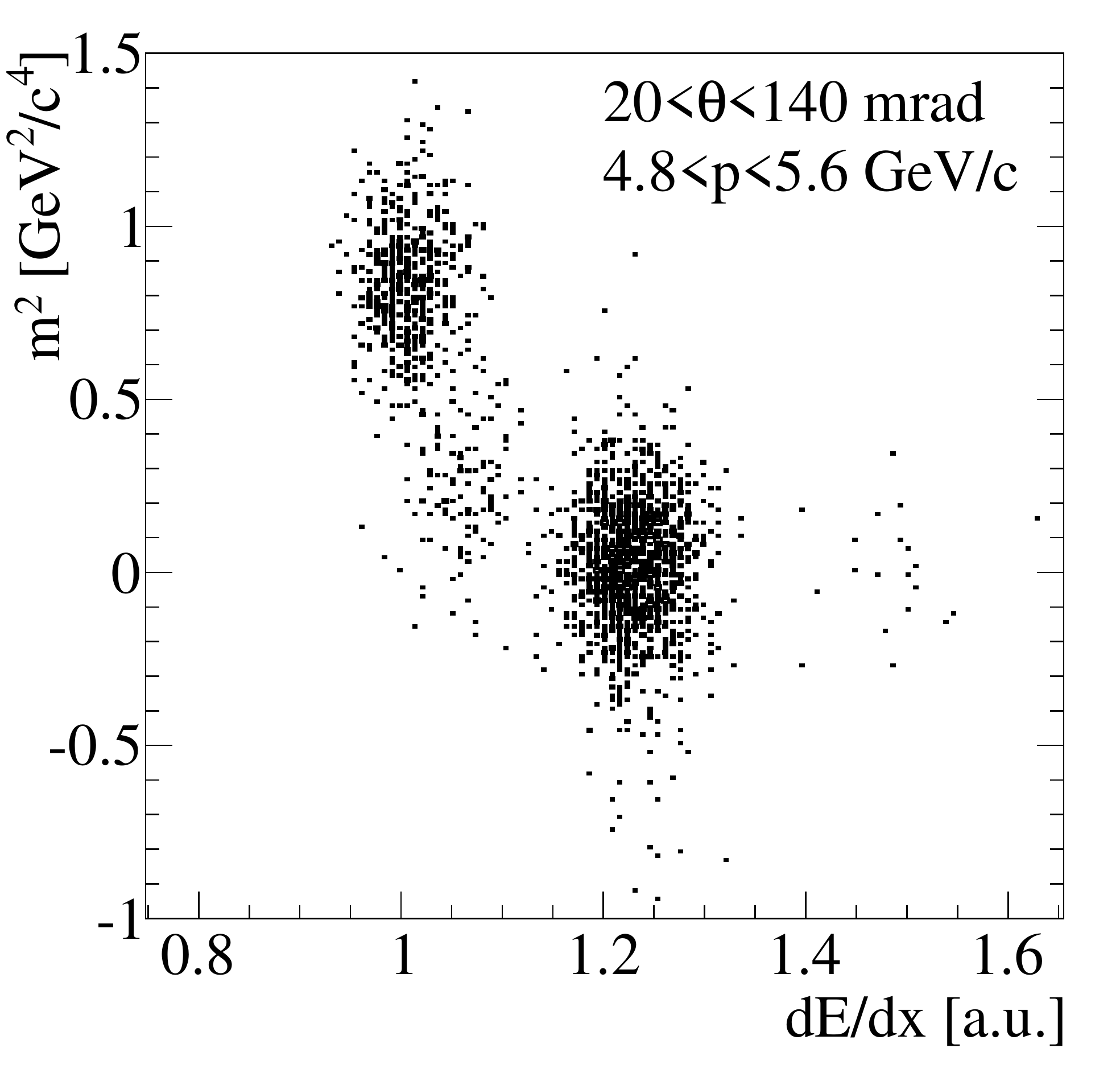}
    \includegraphics[width=0.32\linewidth,scale=0.4]{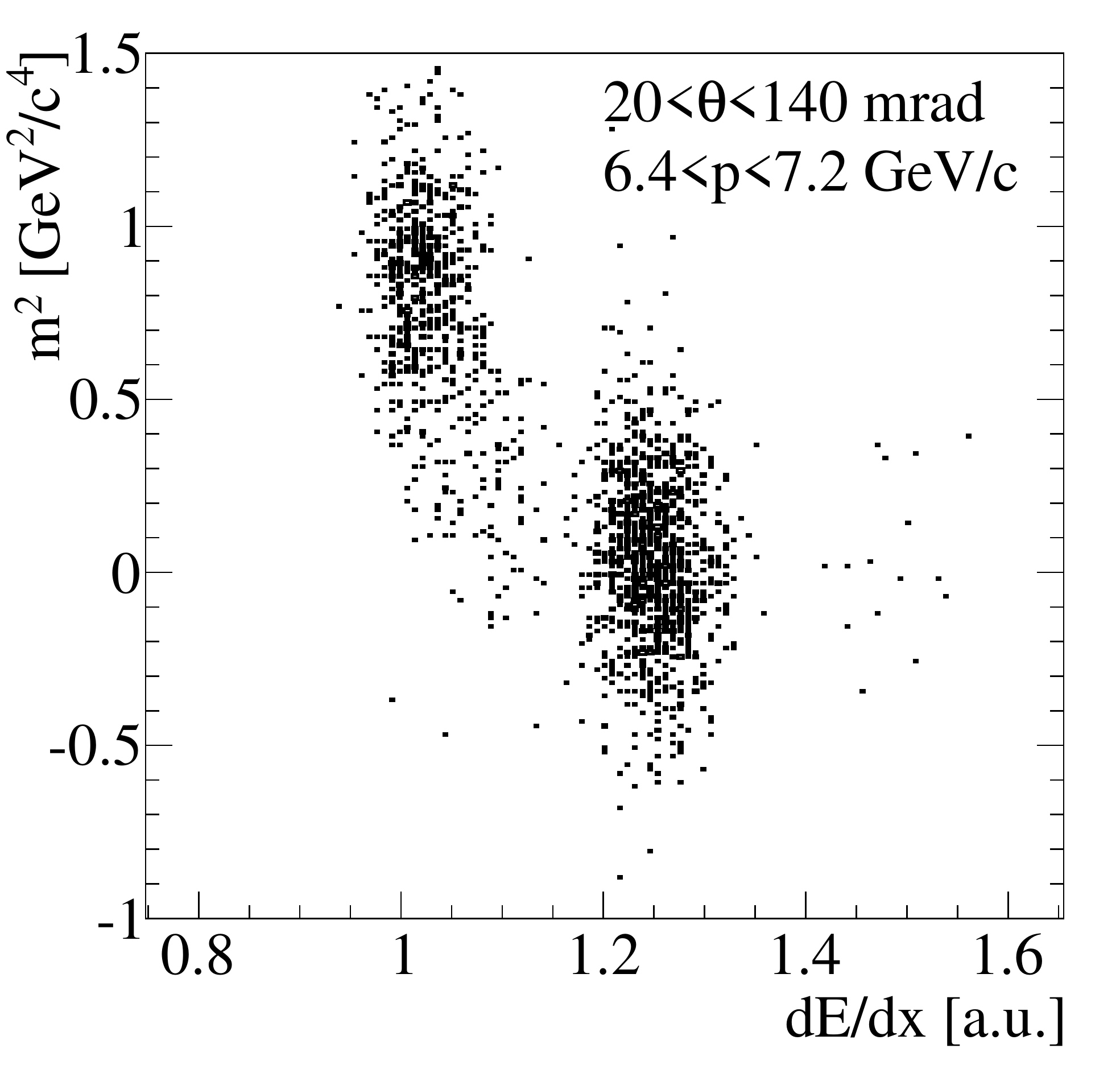}
  \end{center}
    \vspace{-0.3cm}
    \caption{\label{fig:scatter3} 
      %Correlation between the
      Scatter plots of $dE/dx$ versus $m^2$ 
      measured for the selected tracks in three
      $\{p,\theta\}$ bins.}
\end{figure*}
\begin{figure*}[htb]
  \begin{center}
    \includegraphics[width=0.32\linewidth,scale=0.4]{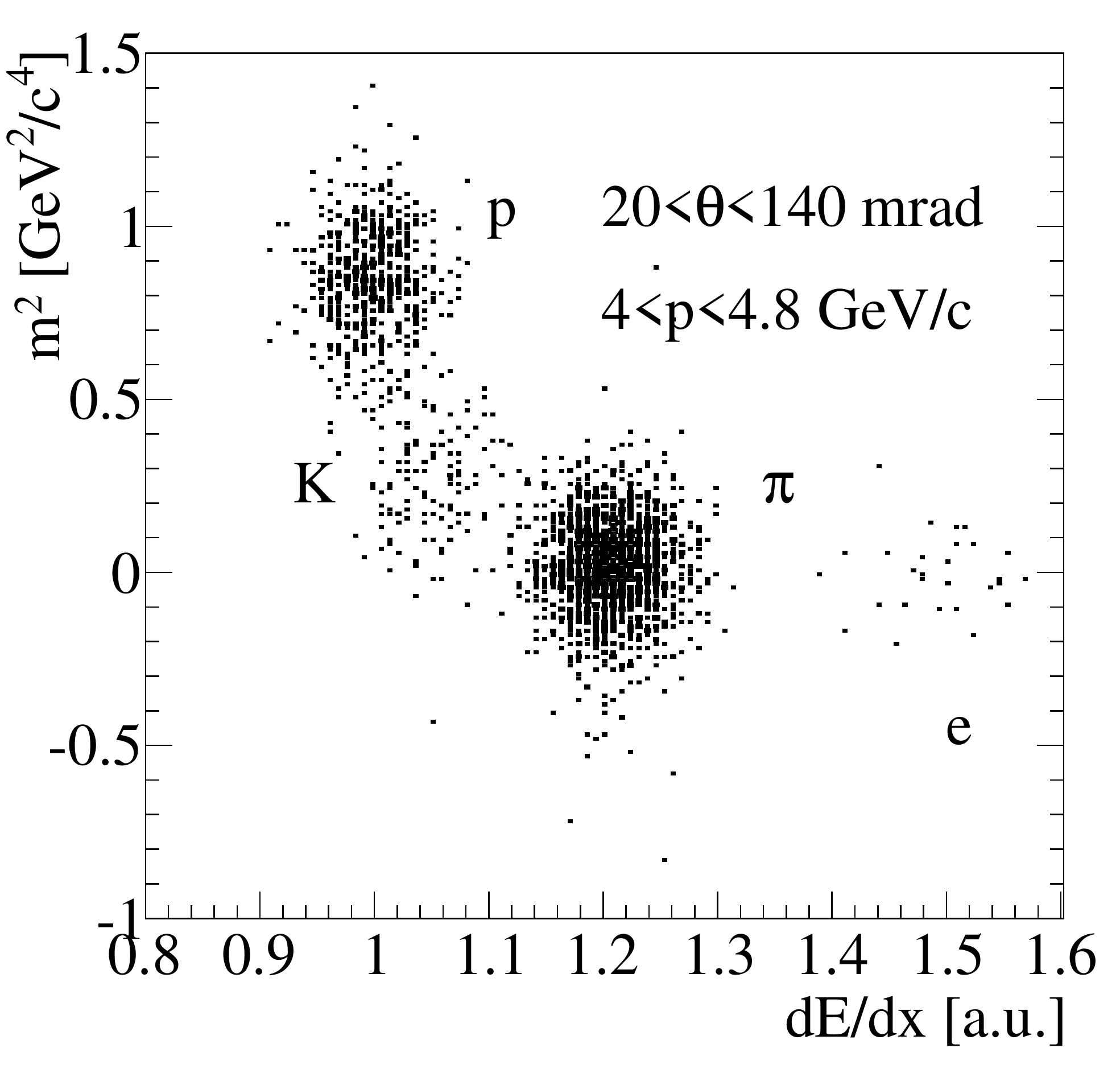}
    \includegraphics[width=0.32\linewidth,scale=0.4]{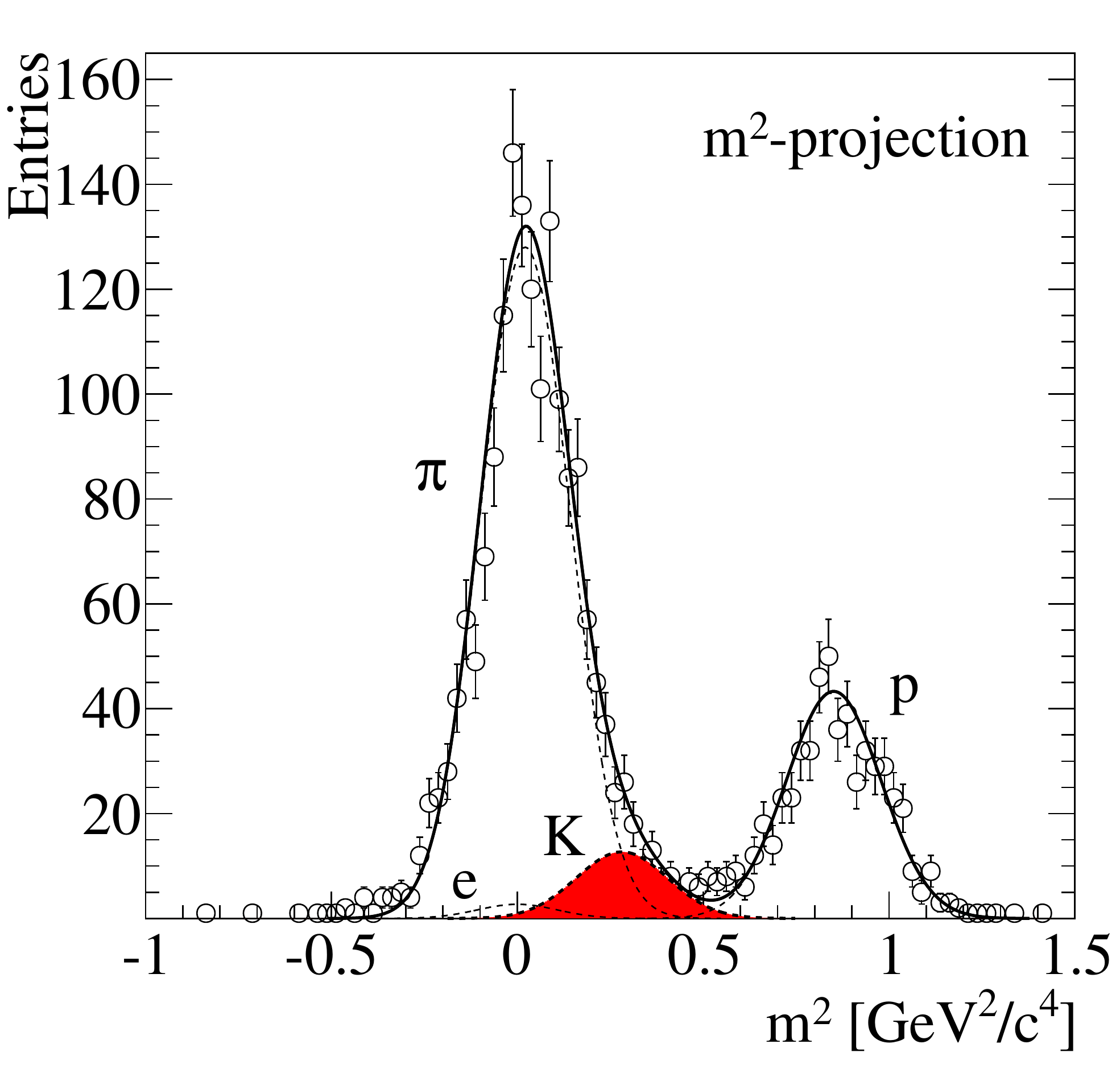}
    \includegraphics[width=0.32\linewidth,scale=0.4]{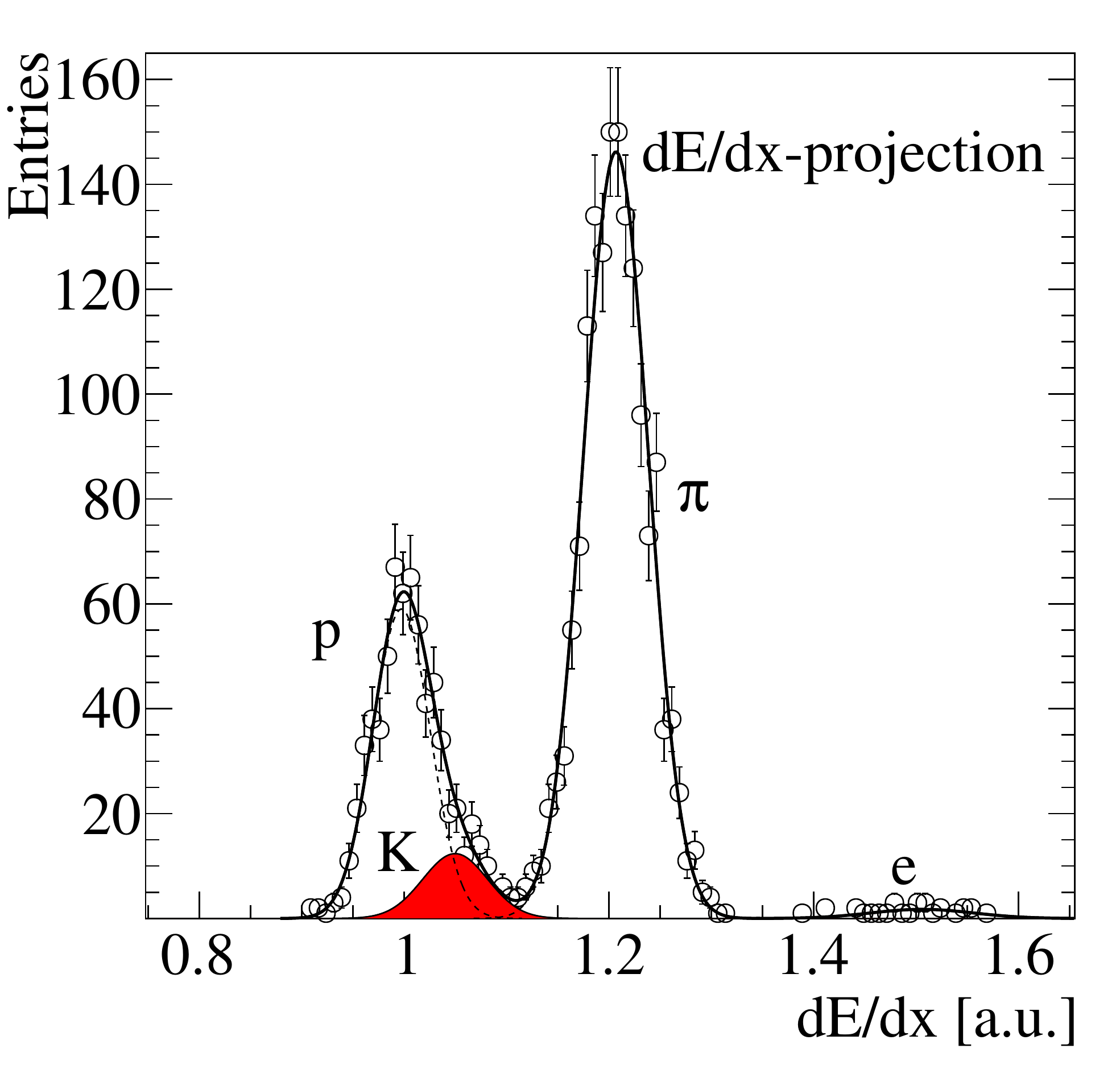}
  \end{center}
  \vspace{-0.3cm}
  \caption{\label{fig:scatter} Example of a bi-dimensional fit to the
    $dE/dx$-$m^2$ distribution. The $m^2$ and $dE/dx$ projections are
    also shown superimposed with the results of the fitted functions.
  }
\end{figure*}
\begin{figure}[]
  \begin{center}
    \includegraphics[scale=0.4]{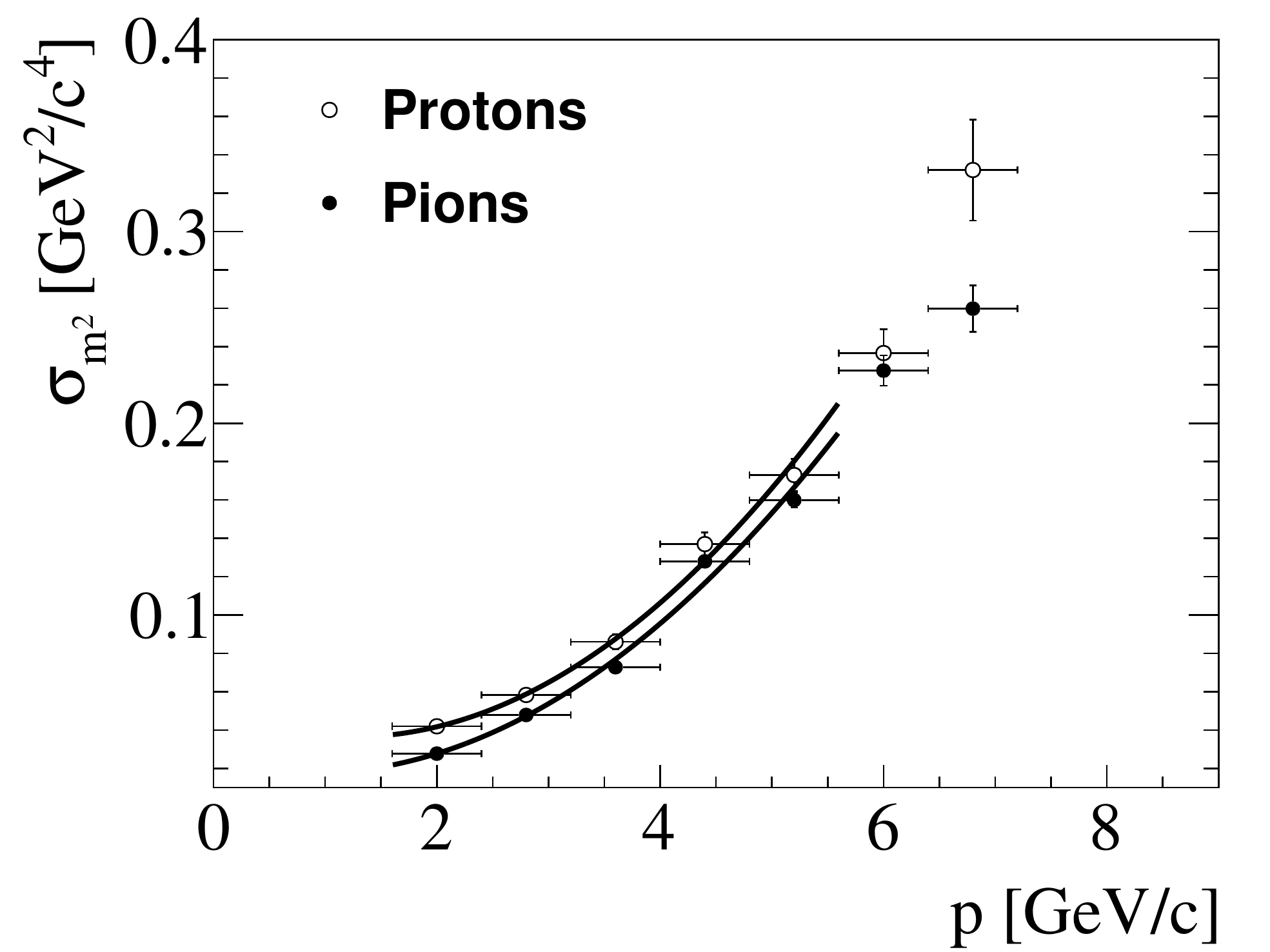}
     \vspace{-0.3cm}
  \end{center}
  \caption{\label{fig:m2res} The $m^2$ resolution versus the measured
    momentum for tracks with 20$<\!\theta\!<$140 mrad.  Parabolic fits
    to the best measured points are superimposed.  }
\end{figure}

\begin{figure}[]
  \begin{center}
    \includegraphics[scale=0.4]{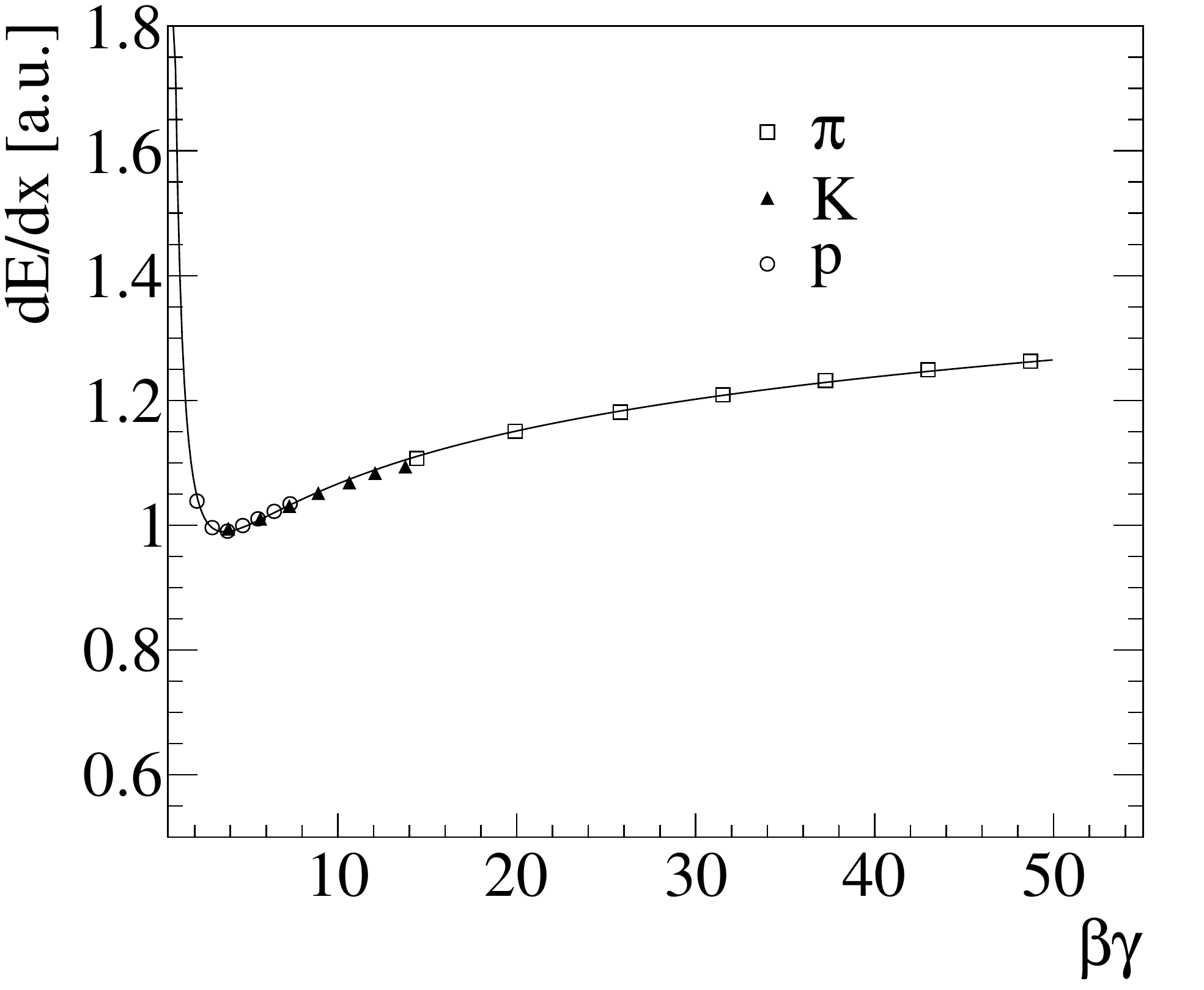}
   \vspace{-0.3cm}
   \caption{\label{fig:bbfit} Bethe-Bloch function fit to mean $dE/dx$
     values (data points). %obtained from data.
     The points used in the fit are from protons and kaons of
     $\beta\gamma<6$ and pions of $\beta\gamma>20$.}
\end{center}
\end{figure}

The specific ionization energy loss, $dE/dx$, measured in the TPCs,
and the time-of-flight, measured with the ToF-F, are used to identify
charged particles of different masses in the chosen momentum
intervals.

For each track, the $dE/dx$ is defined as the mean calculated
from the lowest $60\%$ of the cluster charges.
This reduces the effect of the long Landau tail in the
single cluster charge distribution to the extent that
the $dE/dx$ experimental resolution can be assumed to be
Gaussian within the precision which
can be achieved with the available statistics.

A dedicated calibration procedure is applied to the data to correct
the measured cluster charge deposits for various detector effects,
e.g. charge absorption along the drift path, effective sample length
and variations of gain between the TPC sectors. After the calibration
the corrected data are fitted to a function parameterizing the
Bethe-Bloch relation~\cite{stern}.  This function can then be used to
calculate the mean $dE/dx$ (the mean specific energy loss) for a given
momentum and for a given mass hypothesis.

Kaon identification based only on the $dE/dx$ information is
impossible in the momentum bins below about 4~GeV/$c$, since protons
and kaons have similar values of $dE/dx$ in this kinematic range
(see Fig.~\ref{fig:scatter3}).\\
Another approach is to identify particles through the measurement of
their mass squared $m^2$.  This method is based on the measurement of
the time-of-flight which along with the track length and momentum
allows the calculation of the particle mass.

Identification based only on the $m^2$ information is effective at low momenta, 
below about 4~GeV/$c$, 
where the separation power is higher (Fig.~\ref{fig:scatter3}) than for $dE/dx$.
The uncertainty on the mass measurement is in fact dominated by the time resolution,
meaning that the $m^2$ resolution worsens quadratically with increasing momentum (Fig.~\ref{fig:m2res}).\\

Given the complementarity of the two approaches the particle
identification (PID) capability can be improved, over the whole
kinematic range, by using the combined $dE/dx$ and $m^2$ information.

Figures ~\ref{fig:scatter3} and \ref{fig:scatter} show distributions
of measured $dE/dx$ versus $m^2$ in the first angular interval,
$20<\!\theta<\!140$~mrad for several momentum bins.  Accumulations
corresponding to the different particle types are clearly observable
(in particular for kaons) in contrast to the individual $dE/dx$ and
$m^2$ distributions.

The $dE/dx-m^2$ function used to fit
the yields of
protons ($p$), kaons ($K^+$), pions ($\pi^+$) and positrons ($e^+$)
to the data is
a superposition of four bi-dimensional Gaussians.

A dedicated Monte Carlo simulation was used to validate the accuracy
of the simple Gaussian approximation so as to ensure that no biases
related to the definition of the particle distribution function are
introduced.  The selected function has 20 parameters: 4 yield-, 8
width- and 8 mean-parameters.  In all analyzed bins the $e^+$
accumulations in the $dE/dx$-$m^2$ plane are fully separated from
other particles (Fig.~\ref{fig:scatter3}).  Thus the number of
parameters relevant for the kaon yield determination reduces to 15.
Arguments and details concerning the selected function and further
reduction of the number of fitted parameters are briefly outlined
below.
\begin{itemize}
\setlength{\itemsep}{1pt}
\item[(i)]{  {\it Width of the $dE/dx$ distributions.} \\
    The measured width of the different $dE/dx$ peaks is the result of
    the experimental (detector) resolution plus the smearing due to
    the variation of the $dE/dx$ mean value with the momentum.  The
    second contribution depends on particle type (mass) and is
    non-negligible in the lower momentum bins, where the relativistic
    rise of the energy loss is steep.  In order to
    separate %disentangle
    the two contributions the variance of each Gaussian $\sigma^2_i$
    ($i \equiv p,K,\pi$) was expressed as the sum of two terms:
    $\sigma^2_i=\sigma^2_{exp}+\sigma^2_{i,bin}$.  The first is the
    experimental resolution and is treated as a free fit parameter.
    The second, a constant term, was derived as the variance of the
    distribution of the $dE/dx$ values calculated from the Bethe-Bloch
    function for all the tracks in the respective $\{p,\theta\}$ bin.
    Despite the fact that momentum distributions of protons, kaons and
    pions differ somewhat, the momentum distribution for all particles
    was used to evaluate $\sigma^2_{i,bin}$.  This procedure was
    justified by Monte Carlo studies which showed that, in the
    considered kinematic range, a particle-type-dependent momentum
    distribution has no relevant impact on the outcome of the
    $\sigma^2_{p,bin}$, $\sigma^2_{K,bin}$ and $\sigma^2_{\pi,bin}$
    calculation.
       
         The assumption of a resolution $\sigma_{exp}$ independent of
         the particle type (mass) was validated by investigating the
         dependence of the $dE/dx$ resolution on the mean $dE/dx$.
         From a fit performed to the pion distribution in bins where
         the accumulation is well isolated, no indication of a
         dependence of the peak width on the peak position was found.
         This result allows to use the same experimental resolution
         for all hadron species.

         The experimental resolution is inversely proportional to the
         square root of the number of $dE/dx$ measurements for a
         track.  The latter vary from track to track.  Therefore the
         actual distribution of the energy loss is rather a
         superposition of Gaussians of different widths.  The simple
         Gaussian approximation is still applicable since the topology
         of the selected tracks results in a narrow distribution of
         the number of points peaked above 100 in each $\{p,\theta\}$
         interval.  This finding was validated through a dedicated
         Monte Carlo simulation.  }
     \item[(ii)]{ {\it Mean of the  dE/dx distributions}\\
         Since the mean energy loss depends only on the momentum to
         mass ratio, the $dE/dx$ distributions of different particle
         species with the same momentum distribution are shifted.  In
         the fit only the pion mean energy loss was a free parameter.
         The kaon and proton mean-paramters were instead calculated
         using the fitted pion mean and the shifts calculated from the
         Bethe-Bloch parametrization.  }
\item[(iii)]{ {\it Width of the $m^2$ distributions}\\
The Gaussian approximation is
adopted also for the 
projected $m^2$ distributions. 

The MC simulation shows that the distortion of the Gaussian shape due to
the increase of the $m^2$ variance with the momentum has a negligible impact
on the fitted yields. The $m^2$ width-parameters of $p$'s, $K$'s and $\pi$'s are
fitted independently.
}
\item[(iv)]{ {\it Mean of the $m^2$ distributions}\\
The $m^2$ mean-parameters of $p$'s, $K$'s and $\pi$'s are
fitted independently.
}

\end{itemize}
Thus the fitted parameters relevant for the kaon yield
determination are: three yield-parameters as well as one width- and one
mean-parameter for the $dE/dx$ Gaussians, as well as three width- and
three mean-parameters
for the $m^2$ Gaussians.

Performance studies carried out
on simulated data
show that,
in the higher momentum bins
where the $K/p$ and $K/\pi$ production ratios rapidly decrease
and the peaks significantly overlap,
the results depend strongly on the details of the fit procedure,
e.g. initial values and bounds of the fitted parameters.
This in particular concerns the
$dE/dx$ and $m^2$ mean-parameters as well as the  $m^2$ width-parameters.
Along with the definition of the fit function, it
is therefore mandatory to establish
a procedure for the
precise determination of the initial values and bounds
of the fitted parameters.

For the distributions in $m^2$,
the strategy  is to estimate
mean- and width-parameters at
high momenta by extrapolating
the fit results from low momentum bins.

First, mean-parameters of
protons, kaons and pions
are fitted
in the bins in which particle accumulations are well
separated
(e.g. up to 4~GeV/$c$ for
20$<\!\theta\!<$140~mrad).
The fitted values are independent
of the momentum, which
proves the correctness of the ToF-F calibration procedure.
The average of the fitted  mean-parameters and  its error
are used as the initial value and bounds, respectively, in the final fit.
Second, the $m^2$ width-parameters are fitted over the whole momentum
range.  Figure~\ref{fig:m2res} shows the results for protons and
pions.  As expected, the momentum dependence of the $m^2$
width-parameter is well fitted by a second order polynomial. The
width-parameter is larger for protons than pions, since for the same
momentum and path length, the time of flight is larger for heavier
particles. The kaon width-parameter cannot be fitted because of the
small separation in $m^2$ of the kaon and pion Gaussians.  It is
therefore calculated using the fitted pion width at the momentum bin
center rescaled for the correct mass value.  The proton and pion
resolution functions are derived via parabolic fits extended over the
momentum range where peaks are well separated (e.g. up to 5.6 GeV/$c$
for 20$<\!\theta\!<$140~mrad, see Fig.~\ref{fig:m2res}).  Initial
values for width-parameters are therefore calculated by extrapolating
the resolution functions to the center of the momentum
bin. Extrapolation errors are calculated as well and used to constrain
the allowed fit parameter range.

Figures~\ref{fig:scatter3} and~\ref{fig:scatter} show that in the
$dE/dx$-$m^2$ plane the peak of the pion accumulation can be well
determined in all analyzed bins.  For protons and kaons this is
possible in several bins.  Thus a precise fit of the $dE/dx$
mean-parameter is possible over the whole $\beta\gamma$ range relevant
for the analysis.  The results are presented in~Fig.~\ref{fig:bbfit}.
Points with low and intermediate $\beta\gamma$ correspond to protons
and kaons while points with high $\beta\gamma$ correspond to
pions. Thus, points measured with high precision are the ones with
lower and higher $\beta\gamma$ while the intermediate region
corresponds to protons and kaons of high momenta for which the
accumulations overlap (Fig.~\ref{fig:scatter3}).

This means that, for intermediate $\beta\gamma$ values, an accurate
parametrization of the expected $dE/dx$ can be obtained from a
Bethe-Bloch curve fitted only to the high precision measured points,
which are in the low- and high- $\beta\gamma$ region, as specified in
Fig.~\ref{fig:bbfit}.

The precision of this procedure, estimated from the residuals with
respect to the fit points, turned out to be about 0.5\%.  This value
meets the requirement, derived from the Monte Carlo studies, of a
precision below 1\% needed to keep the systematic error below the
statistical uncertainty.  Estimates of the systematic errors are
reported in Sec.~\ref{Sec:sys}.  It is worth noting that the required
precision on the determination of the $dE/dx$ mean-parameters cannot
be achieved by using the Bethe-Bloch curve as calculated from the
calibration procedure since this represents an overall parametrization
of data averaged over the detector as a whole covering a wider range
of track topologies.

Finally, an example of the function fitted to data using
the Maximum Likelihood method is shown in Fig.~\ref{fig:scatter}.
\subsection{Correction factors}\label{Sec:corrections}
\begin{figure}[htb]
  \begin{center}
    \includegraphics[scale=0.42]{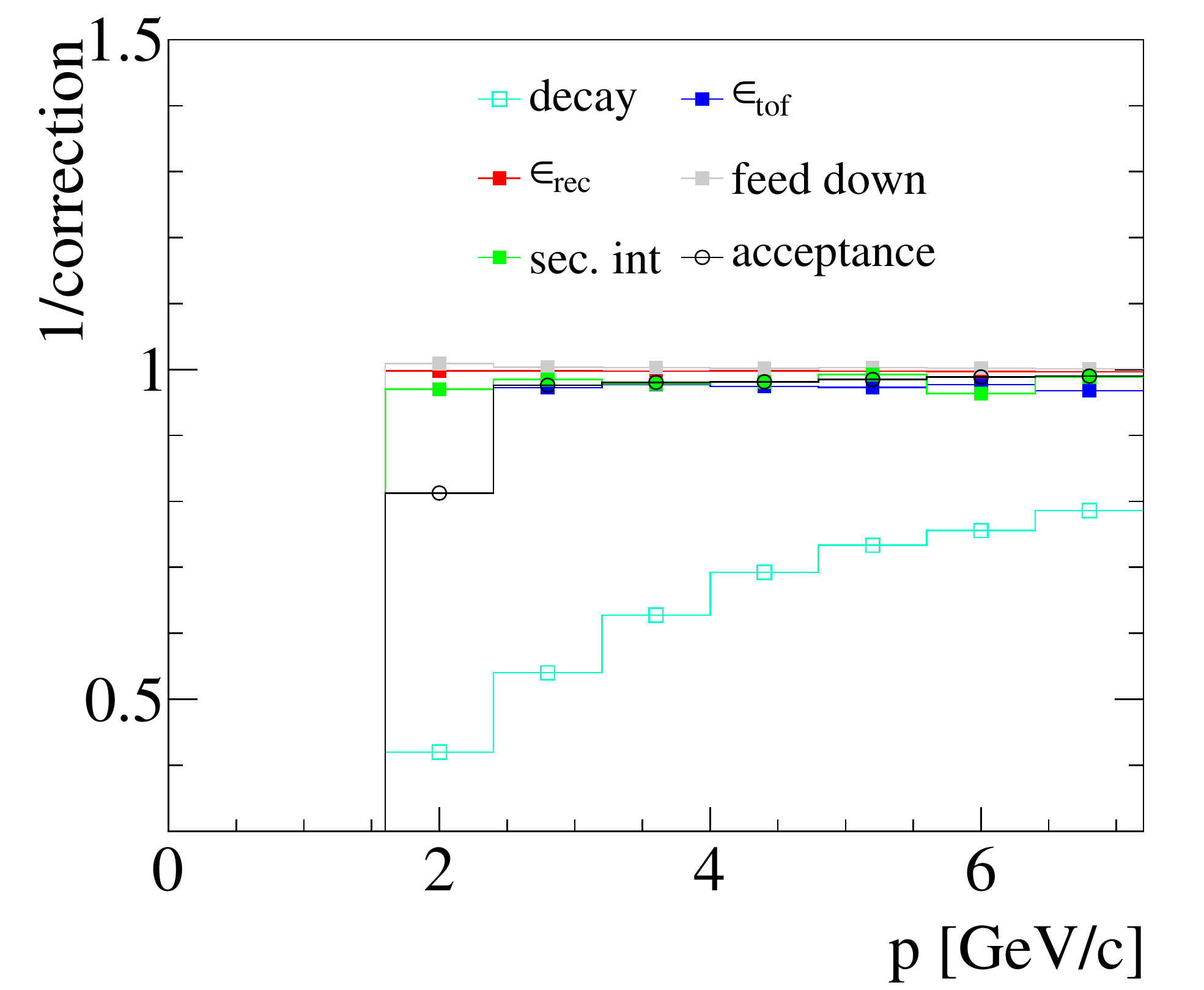}
  \end{center}
  \vspace*{-0.5cm}
  \caption{(Color online) Monte Carlo correction factors as a function
    of momentum calculated for the angular interval $20<\theta<140$
    mrad.}
  \label{fig:cor_seperate}
\end{figure}

The Monte Carlo simulation described in Refs.~\cite{pion_paper}-Sec. IVC 
and~\cite{Nicolas}
was used to calculate corrections for kaon decay, secondary
interactions in the target and detector material and track reconstruction
efficiency. Two strategies were implemented to correct the raw
spectra:
\begin{itemize}
\item [(i)]{the different biases were calculated
  separately with the MC and applied successively to the data,}
\item [(ii)]{a global Monte Carlo correction taking into account all the
    above effects.}
\end{itemize}
The MC correction factors calculated with the first
strategy  are shown in Fig.~\ref{fig:cor_seperate} for
the first angular bin. The separation of corrections clearly shows
that the decay correction dominates.
This correction was estimated, for each $\{p,\theta\}$ bin, by computing the
 fraction of kaons produced in the primary interaction
which reach the ToF-F wall before decaying.
As such, this procedure assumes that the track selection defined
in Sec. \ref{sec:track_sel}, in particular the cut on the $z$
position of the last measured point ($z_{last}$), is fully efficient in the
isolation of a pure sample of stable kaons.
Before applying this correction it is therefore
necessary to correct
the sample of identified kaons
for the contamination
from those kaons which decayed
before reaching the ToF-F but were still
associated to a ToF-F hit (see Sec. \ref{sec:track_sel} item (v) and (vi)).
This contamination was found to be of about 2-3$\%$.

The feed down correction
concerns kaon tracks fitted to the primary vertex but not produced in the
primary interaction. Feed down is only observed from kaons produced by
secondary interactions in the target and the effect is below 2$\%$ over
the whole kinematic range.

The ToF-F efficiency was accounted for and calculated from the data by
requiring that a track traversing the ToF-F wall generates a hit in
the ToF-F.

Except for the first momentum bin, the small correction for geometrical
acceptance 
(on top of the factor due to the selection of the $\phi$ wedge)
reflects the fact that only full acceptance regions were
selected as described in Sec.~\ref{sec:track_sel}. Tracks are also
required to hit ToF-F. For these long tracks the reconstruction efficiency
is close to unity.

The contribution related to kaon losses due to secondary interactions in
the detector material is about 2\%.

%\\
Within the second strategy,
global correction factors were calculated by means of
different detector response unfolding  methods \cite{unfolding}.
All the applied methods returned results whose discrepancies are
significantly below the associated statistical errors.

Correction factors obtained through the unfolding technique were
compared to those calculated as the product of the individual factors
derived from method (i).  Results turned out to be compatible within
the statistical fluctuations and no systematic biases were observed.
This implies that all the relevant contributions to the corrections
were identified and quantified correctly and that the different
contributions are weakly correlated.  Moreover, since the unfolding
procedure accounts for event migration between adjacent bins, the
compatibility of the two approaches proves that bin-by-bin correlation
of the correction factors is negligible as well.

\subsection{Systematic errors}\label{Sec:sys}
The main contribution to the overall systematic error on the
yield correction arises from the decay correction.

Three main contributions to the systematic uncertainty of the decay correction
are discussed below.
\begin{itemize}
\item[(i)] {\it The contribution due to kaons decaying in flight.} \\
  Most of the unstable kaons in the selected final sample are decays
  which take place in the region of space delimited by $z_{last}$ and
  the ToF-F surface.  An alternative way to derive the decay
  correction is then to calculate the survival probability only until
  $z=z_{last}$.  In this procedure one must correct for the small
  contamination of kaons decaying between $z_{last}$ and the ToF-F
  which are not associated to a ToF-F hit. The two methods give
  results which differ by less than $1\%$.  The systematic error
  assigned to the bias related to kaon decays was therefore
  conservatively calculated as $50\%$ of the corresponding
  contamination (see Sec. \ref{Sec:corrections}).
 \item[(ii)] {\it The contribution due to the model dependence of momentum distributions.} \\
   The value of the decay correction depends on the momentum
   distribution within each $\{p,\theta\}$ bin and therefore on the
   specific MC model used for the event generation.  Significant
   discrepancies may be observed especially at low momenta where the
   decay probability (the correction) is larger and the momentum
   distribution is steeper.  This effect was quantified by using
   various MC models and comparing the resulting correction factors
   with those calculated directly from the data.  More precisely, the
   latter factor was derived for the three lowest momentum bins, where
   unique identification of kaons is possible on a track-by-track
   basis: the raw yield was corrected by re-weighting each track with
   its decay probability, calculated for the measured momentum and
   track length.  The resulting deviation of the correction factor was
   negligible except in the first momentum bin where a maximum
   difference of about 2\% was found.

\item [(iii)] {\it The contribution due to the uncertainty in the reconstructed track
  length.} \\
Since only well measured tracks which traverse the entire
spectrometer are retained, a precision of a few millimeters is achieved
on both track length calculation
and extrapolation to the ToF-F surface.
This translates into a negligible (order of 0.1~\%)
error on the decay correction factor.
\end{itemize}

Other sources of systematic errors include uncertainties on
the ToF-F and reconstruction efficiency and on the contribution
from secondary interactions.

Systematic uncertainties on the ToF-F efficiency come from the accuracy of the calibration
procedure. The estimated value is 2$\%$.

The systematic error on the
reconstruction efficiency was estimated by varying the track
selection cuts. The induced bias is small compared to the statistical fluctuations;
this is expected considering the high reconstruction efficiency.

Finally a systematic error corresponding to 30$\%$ of the correction was assigned to the
contribution of secondary interactions \cite{pion_paper}.\\
\begin{figure}[htb]
  \begin{center}
    \includegraphics[scale=0.4]{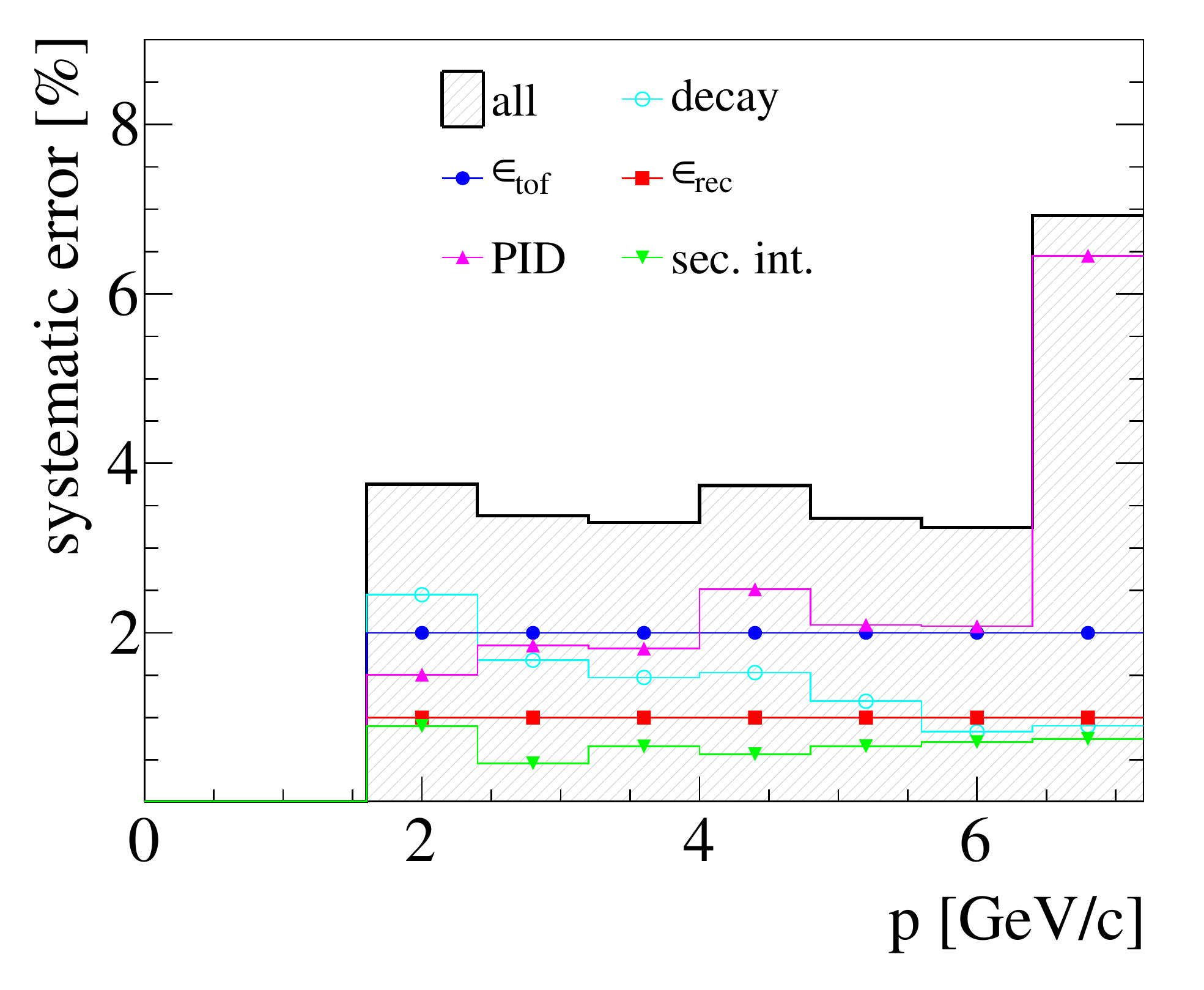}
  \end{center}
  \vspace{-0.5cm}
  \caption{(Color online) Breakdown of systematic errors as a function
    of momentum for the angular interval \mbox{$20<\theta<140$ mrad.}}
  \label{fig:sys}
\end{figure}
\newline
Systematic uncertainties related to the procedure used for the PID
(Sec.~\ref{Sec:combined}) were
quantified by studying
the dependence of the fitted kaon yields on the input fit
parameters, namely: the central values of the
$dE/dx$ and $m^2$ distributions and the widths in $m^2$.
The
systematic error was obtained by varying both the input
parameters and limits
and considering the subsequent variation of the returned particle yield.

In particular, the relative distances of the proton, kaon and pion
$dE/dx$ peaks were varied by an amount corresponding to the largest of
the residuals between the fitted Bethe-Bloch curve and the measured
points (see Sec.~\ref{Sec:combined}).  The allowed range for the $m^2$
peak position and resolution were enlarged by a factor of three with
respect to the original value which had been derived from the errors
in the parametrization as described in Sec.~\ref{Sec:combined}.
Results are shown in Fig.~\ref{fig:sys}.

As underlined in the introduction of the Sec.~\ref{Sec:ana}, the sensitivity to the
kaon signal decreases with the momentum,
therefore, the fitted kaon yield depends significantly
on the definition of the input parameters.
This explains the steep increase of the systematic error in the last momentum bin.

The relative contributions of all considered systematic errors are
shown in Fig.~\ref{fig:sys}
for the first angular interval.\\

\section{Results}

The $K^+$ spectra presented in this paper refer to positively
charged kaons produced in strong and electromagnetic processes in p+C
interactions at 31 GeV/$c$.
 Differential inclusive $K^+$  cross sections were derived following 
the procedure described in \cite{pion_paper,claudia}.

\begin{figure}[ht]
  \begin{center}
    \includegraphics[scale=0.43]{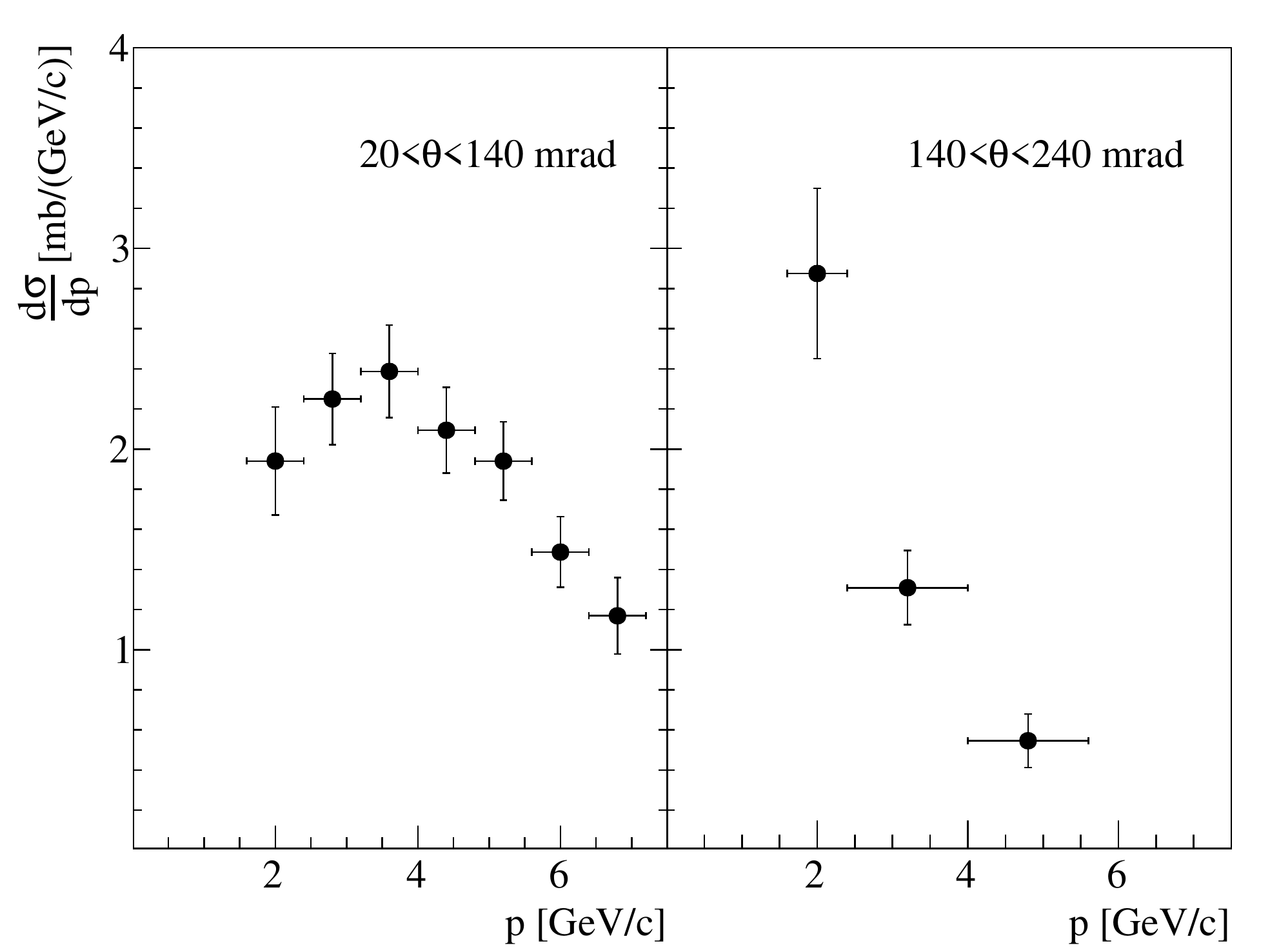}
    \vspace{-0.3cm}
    \caption{Differential cross sections for $K^+$ production in p+C
      interactions at 31 GeV/$c$. The spectra are presented as a
      function of laboratory momentum, $p$, in two intervals of polar
      angle, $\theta$. Error bars indicate statistical and systematic
      uncertainties added in quadrature. The overall uncertainty
      (2.5\%) due to the normalization procedure is not included.}
  \label{fig:k_results}
  \end{center}
\end{figure}

The results are presented in Fig.~\ref{fig:k_results} and
Table~\ref{tab:k_results} as a function of momentum in the two considered
intervals of polar angle. Momentum and polar angle are calculated in the
laboratory system.

\begin{figure}[ht]
  \begin{center}
    \includegraphics[scale=0.43]{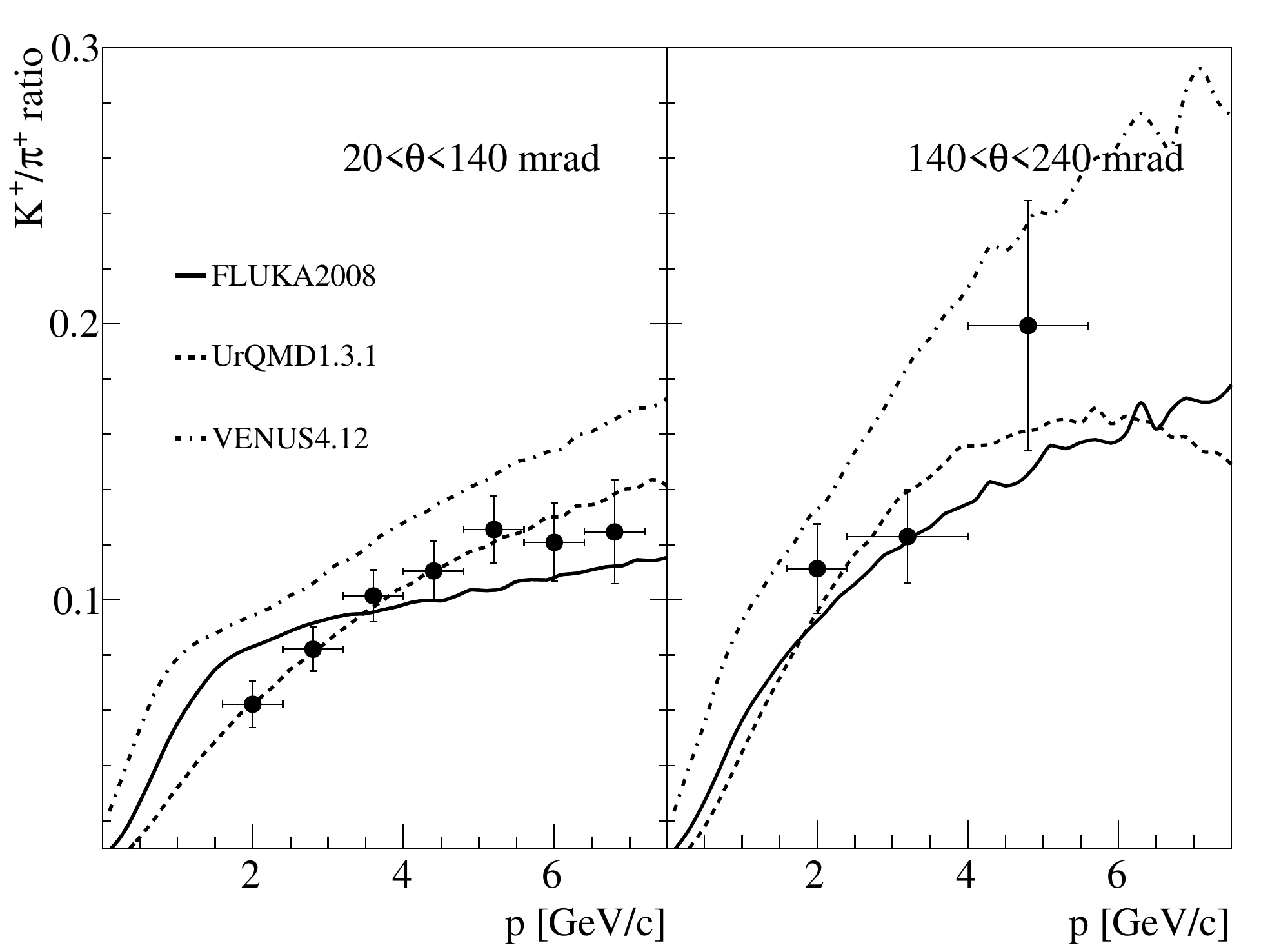}
  \end{center}
   \vspace{-0.3cm}
   \caption{Ratio of $K^+$ to $\pi^+$ production cross sections in p+C
     interactions at 31 GeV/$c$. The ratios are presented as a
     function of laboratory momentum, $p$, in two intervals of polar
     angle $\theta$. Errors are calculated taking into account only
     statistical uncertainties.  Predictions of hadron production
     models are superimposed.  }
  \label{fig:kpiratio}
\end{figure}

The ratio of $K^+$ to $\pi^+$ production cross sections is shown in
Fig.~\ref{fig:kpiratio}.
The $\pi^+$ spectra are taken from Ref.~\cite{pion_paper}, 
central values and errors are recalculated to match the binning
used in the current analysis.
Numerical values are reported in Table~\ref{tab:k_results}.

\begin{table*}[!ht]

\renewcommand{\tabcolsep}{3mm}
\begin{tabular}{rr|rr|r|r|ccc|ccc}
\hline
$\theta_{low}$ &
$\theta_{high}$ &
$p_{low}$ &
$p_{high}$ &
$N$ &
$N_K$ &
$d\sigma^{K^+}\!/dp$ &
$\Delta_{stat}$ &
$\Delta_{sys}$&
$\sigma^{K^+}\!\!/\sigma^{\pi^+}$ &
$\Delta_{stat}$
\\
\multicolumn{2}{c|}{(mrad)}&
%\multicolumn{1}{c}{mm}&&
\multicolumn{2}{c|}{(GeV/$c$)}
&
&
&
\multicolumn{3}{c|}{(mb/(GeV/$c$))}&
\multicolumn{2}{c}{ }
\\
\hline
% &  2395 &  2934 &  2662 & 2263 &  1964 &   1699  &  1424       \\
20 &140 &1.6 &2.4  &2395&56 &1.94 &0.26 &0.07 &0.062 & 0.008 \\

& &2.4 &3.2        &2934 &106 &2.25 &0.21 &0.08&0.082 & 0.008 \\

& &3.2 &4           &2662&134 &2.39 &0.22 &0.08&0.10 &0.01 \\

& &4 &4.8           &2263   &127 &2.10 &0.20 &0.08&0.11 &0.01 \\ %2.09

& &4.8 &5.6         &1964    &126 &1.94 &0.18 &0.07&0.13 &0.01 \\

& &5.6 &6.4         &1699     &97 &1.49 &0.17 &0.05&0.12 &0.01 \\

& &6.4 &7.2         &1424   &81 &1.17 &0.17 &0.08&0.13 &0.02 \\

\hline

140 &240 &1.6 &2.4    & 1399  &49 &2.89 &0.41 &0.11&0.11 &0.02 \\ %2.87
& &2.4 &4            &1340   &64 &1.32 &0.17 &0.06&0.12 &0.02 \\ %1.31
& &4 &5.6            &529   &32 &0.55 &0.12 &0.06&0.20 &0.04 \\

\hline
\end{tabular}
\caption{\label{tab:k_results} The NA61/SHINE results for the differential $K^+$ production
  cross section in the laboratory system, $d\sigma/dp$, and  the $K^+$ to
  $\pi^+$ ratio of production cross sections, $\sigma^{K^+}\!\!/\sigma^{\pi^+}$, for p+C
  interactions at 31 GeV/$c$.
  Each row refers to a different ($p_{low}\!\leq p < p_{up},
  \theta_{low}\!\leq \theta < \theta_{up}$) bin, 
  where $p$ and $\theta$ are the kaon momentum and polar
  angle in the laboratory frame.
  $N$ is the total number of selected tracks and $N_K$ is the fitted raw number of kaons.
  The central value as well
  as the statistical ($\Delta_{stat}$) and systematic ($\Delta_{sys}$) errors
  of the cross section are given. 
  The overall uncertainty (2.5\%) due to the normalization procedure is not included.
  For the  $K^+$/$\pi^+$ ratio, errors are calculated taking into account 
  only statistical uncertainties.
  }
\end{table*}

\section{Comparison with model predictions}

In this section we compare the $K^+$ spectra in p+C interactions at
31~GeV/c with the predictions of hadronic event generators. Three
models, \VenusLong~\cite{Venus1,Venus2}, \FlukaLong~\cite{Fluka}, and
\UrqmdLong~\cite{Urqmd,Urqmd_improve} were selected for this
purpose. They are part of the \Corsika~\cite{Corsika} framework and
are commonly used for the simulation of hadronic interactions at
energies below 80~GeV in extensive air
showers~\cite{model_dependence}. \Venus{} is also the standard
model for acceptance simulations of the NA49 and NA61
Collaborations.

These models were already compared to the pion spectra measured by the
NA61/SHINE Collaboration~\cite{pion_paper}. Motivated by this comparison, a
correction of a technical shortcoming of the \Urqmd{} model was
proposed in~\cite{Urqmd_improve}.  This correction does not affect the
kaon spectra presented here.  Therefore we compare the data with the
original implementation.

\begin{figure}[ht]
  \begin{center}
   
   \includegraphics[scale=0.43,width=1.05\linewidth,height=0.3\textheight]{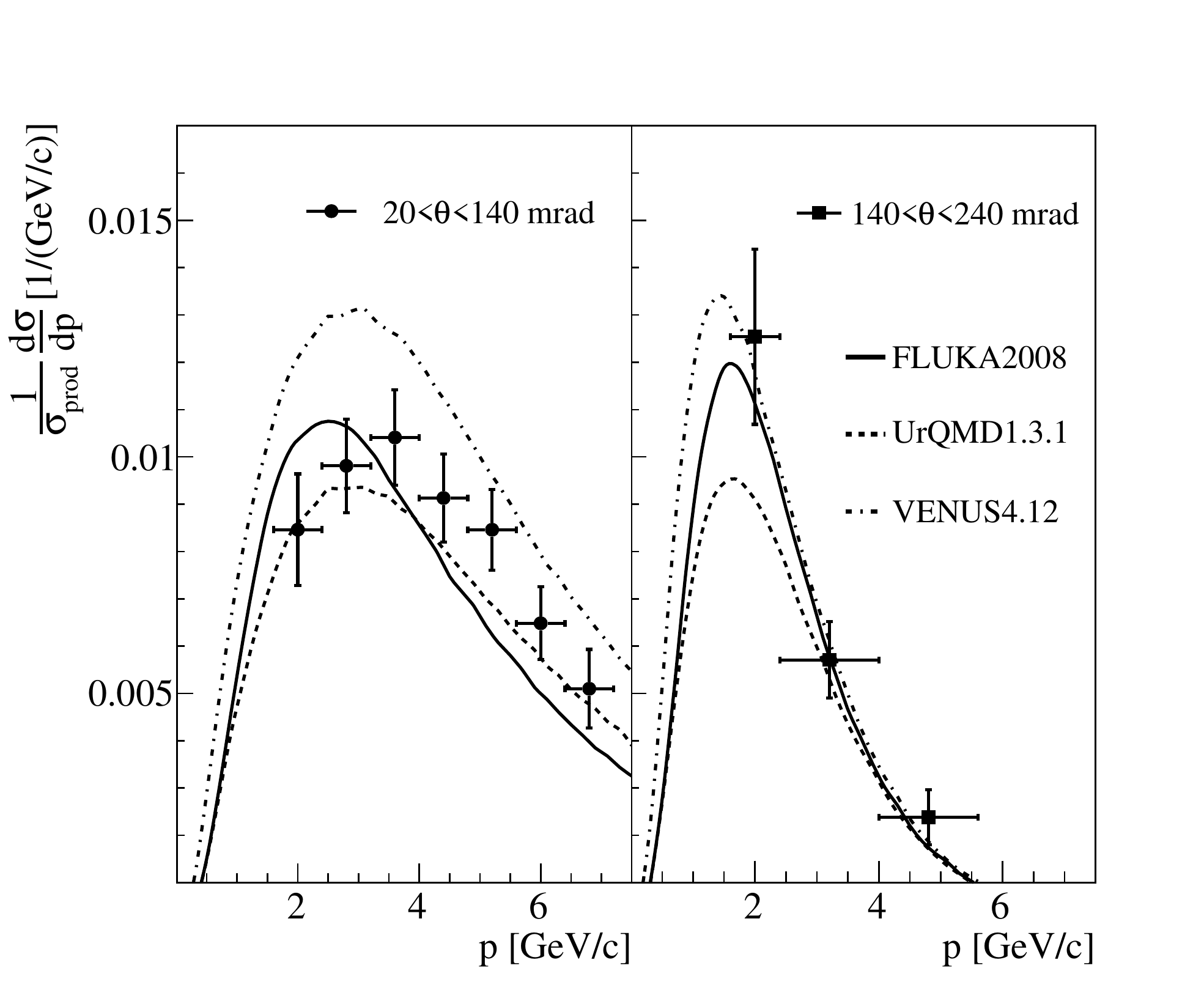}
   \vspace{-0.3cm}
   \caption{Comparison of measured $K^+$ spectra with model
     predictions.  Distributions are normalized to the mean $K^{+}$
     multiplicity in all production p+C interactions.  The vertical
     error bars on the data points show the total (stat.\ and syst.)
     uncertainty. The horizontal error bars indicate the bin size in
     momentum.  }
  \label{fig:k_models}
  \end{center}
\end{figure}

The results of the comparison between data and models are presented in
Fig.~\ref{fig:k_models}. 
In order to avoid uncertainties related to the different treatment of quasi-elastic
interaction and to the absence of predictions for inclusive cross sections,
spectra are normalized to the mean $K^+$ multiplicity in all production interactions.
For the data, the normalization relies on 
the p+C inclusive
production cross section $\sigma_{prod}$ which
was found to be
229.3$\pm$1.9$\pm$9.0~mb.
The production cross section is calculated
from the inelastic cross section by subtracting the quasi-elastic
contribution. Therefore production processes are defined as those in
which only new hadrons are present in the final state. Details of the
cross section analysis procedure can be found in
\cite{pion_paper,claudia}.

The qualitative behaviour of the data is well
reproduced by all models.  The quantitative differences can be related
to the two main production processes of kaons: pairwise production of
a $K^+$ together with another $K$ meson and production of a $K^+$
together with a $\Lambda$ baryon. The latter process dominates kaon
production at large momenta and small angles due to the leading
particle effect.  $K^+$ at large angles and low momenta stem from pair
production of $K$ mesons. Both \Fluka{} and \Venus{} provide a
reasonable description of the pair-produced kaons.  On the other hand,
none of the models is in full agreement with the small-angle
data. While the \Venus{} model overestimates the production of $K^+$
at small angles, \Fluka{} and \Urqmd{} predict a slightly lower kaon
production rate.

Figure~\ref{fig:kpiratio} also shows the comparison between models and data for the ratio of $K^+$ to $\pi^+$
production cross sections:
\Urqmd{} is in good agreement with the data, 
\Fluka{} provides a reasonable description, while \Venus{} 
overestimates the production cross section ratio for both small and 
large angle intervals.

\vspace{-0.3cm}

\section{Summary}

This work presents measurements of differential production cross
sections of positively charged kaons in p+C interactions at
31~GeV/$c$.  These data are essential for precise predictions of the
high energy tail and intrinsic electron neutrino component of the
initial neutrino flux for the T2K long baseline neutrino oscillation
experiment in Japan. Furthermore, they provide important input to
improve hadron production models needed for the interpretation of air
showers initiated by ultra high energy cosmic particles.  The
measurements were performed with the large acceptance NA61/SHINE
spectrometer at the CERN SPS.  A set of data collected with a
4\%~$\lambda_{\mathrm{I}}$ isotropic graphite target during the pilot
NA61/SHINE run in 2007 was used for the analysis.  Positively charged
kaon spectra as a function of laboratory momentum in two intervals of
the polar angle were obtained.  The final spectra were compared with
predictions of hadron production models. The data presented in this
paper are already provided to T2K for the calculation of the neutrino
flux. Meanwhile, a much larger data set with both the thin
(4\%~$\lambda_{\mathrm{I}}$) and the T2K replica carbon targets was
recorded in 2009 and 2010 and is presently being analyzed.

\section{Acknowledgments}
This work was supported by the following funding agencies: the
Hungarian Scientific Research Fund (grants OTKA 68506, 77719, 77815 and 79840), the
Polish Ministry of Science and Higher Education (grants
667/N-CERN/2010/0, N N202 1267 36, N N202 287838 (PBP
2878/B/H03/2010/38), DWM/57/T2K/2007), the Federal Agency of Education
of the Ministry of Education and Science of the Russian Federation
(grant RNP 2.2.2.2.1547), the Russian Academy of Sciences and the
Russian Foundation for Basic Research (grants 08-02-00018 and
09-02-00664), the Ministry of Education, Culture, Sports, Science and
Technology, Japan, Grant-in-Aid for Scientific Research (grants
18071005, 19034011, 19740162, 20740160 and 20039012), the Toshiko
Yuasa Lab.  (France-Japan Particle Physics Laboratory), the Institut
National de Physique Nucl\'eaire et Physique des Particules (IN2P3,
France), the German Research Foundation (grant GA 1480/2-1), the Swiss
National Science Foundation (Investigator-Driven projects and
SINERGIA) and the Swiss State Secretariat for Education and Research
(FORCE grants). The Foundation for Polish Science - MPD program,
co-financed by the European Union within the European
Regional Development Fund.\\
%Swiss Nationalfonds Foundation (grant 200020-117913/1)
%and ETH Research Grant TH-01 07-3.
The authors also wish to acknowledge the support provided
by the collaborating institutions, in particular,
%(for the Swiss) -->
the ETH Zurich (Research Grant TH-01 07-3),
the University of Bern and the University of Geneva.\\
Finally, it is a pleasure to thank
the European Organization for Nuclear Research
for a strong support and hospitality
and, in particular, the operating crews of the CERN SPS accelerator
and beam lines who made the measurements possible.

%\clearpage
%\addcontentsline{toc}{chapter}{Bibliography}
\bibliographystyle{apsrev} 

\bibliography{bibliography}

\end{document}